\documentclass[12pt]{article}
\usepackage{amsmath, amssymb}
\usepackage{authblk}
\title{Comments on ``The impact of Solar magnetic field configurations on the production of gamma rays at the Solar disk'' (arXiv:2512.01403)}

\author{M. Nicola Mazziotta\thanks{mazziotta@ba.infn.it}}

\affil{Istituto Nazionale di Fisica Nucleare, Sezione di Bari, via Orabona 4, Bari, I-70126, Italy}

\date{}

\begin{document}
\maketitle

\begin{abstract}
In this short comment, I discuss the relationship between the results presented in arXiv:2512.01403 and those previously published in Phys.~Rev.~D~101,~083011~(2020). The 2020 study provides a full Monte Carlo simulation of cosmic-ray interactions with the solar atmosphere using the FLUKA code, including realistic solar-atmosphere models, PFSS/Parker/BIFROST magnetic-field configurations, and predictions for gamma rays, electrons, positrons, neutrons, and neutrinos. Given the substantial scientific overlap --- particularly in the modelling of hadronic interactions, magnetic-field effects, cascade development, and comparison with Fermi-LAT observations --- a direct comparison is relevant to assess consistency and complementarity. Here I summarize the main points of agreement, highlight differences in modeling assumptions, and outline how the two approaches can jointly contribute to understanding high-energy emission from the solar disk.
\end{abstract}

\section{Introduction}
The recent preprint arXiv:2512.01403~\cite{Dorner:2025ait} presents simulations of gamma-ray emission from Galactic cosmic-ray (GCR) interactions with the solar atmosphere using CRPropa and a set of magnetic-field configurations. This topic overlaps extensively with the analysis in Mazziotta et al.\ (Phys.~Rev.~D~101, 083011, 2020)~\cite{Mazziotta:2020uey}, where a dedicated FLUKA-based framework was developed to model hadronic and electromagnetic cascades in the solar atmosphere, using realistic density, temperature, and composition profiles, and multiple magnetic-field models. Despite the methodological and scientific overlap, the 2020 study is not cited or discussed. In the following, I briefly summarize its main components and compare the two approaches.

\section{Summary of PRD~101,~083011~(2020)}
The 2020 Phys.\ Rev.\ D study provides a comprehensive FLUKA-based Monte Carlo simulation of cosmic-ray interactions with the solar atmosphere. It includes realistic solar-atmosphere models (SSM gs98, Model~S), a detailed shell-based density and composition structure, and several magnetic-field configurations (PFSS, Parker spiral, BIFROST). The framework predicts secondary gamma rays, electrons, positrons, neutrons, and neutrinos from 100~keV to multi-TeV energies, and compares the results with Fermi--LAT observations of the disk component.

In addition, the study presents the expected gamma-ray flux as a function of angular distance from the solar center, yielding radial profiles for the disk component, as well as a HEALPix map of the disk emission. These results provide spatially resolved predictions directly comparable with Fermi--LAT measurements.

Crucially, the 2020 work also delivers the first predictions of the gamma-ray intensity above 100~MeV, 1~GeV, and 10~GeV for emission from the solar disk. These energy-integrated fluxes, obtained by folding FLUKA yields with realistic cosmic-ray spectra and magnetic configurations, were later also discussed in the recent Fermi--LAT solar-cycle analysis published in Astrophys.\ J.\ Lett.\ 989 (2025) 1, L16. These elements make the 2020 framework an essential reference for any study of cosmic-ray interactions near the solar surface.

\section{Comparison with arXiv:2512.01403}
A comparison between the two studies shows clear scientific overlap:

\subsection{Hadronic Interaction Modelling}
FLUKA provides a full treatment of the hadronic and electromagnetic cascade, including heavy nuclei and multi-step interactions. In contrast, the CRPropa-based simulations use simplified cross-section models. This difference may affect the depth and structure of the resulting cascades and the expected gamma-ray yields.

\subsection{Magnetic Field Modelling}
Both studies use PFSS magnetic-field configurations. Mazziotta et al.\ (2020) used multiple Carrington Rotations (CR~2111, 2125, 2138, 2152) to sample solar-cycle variability, based on HMI/JSOC synoptic magnetograms and a source surface at $R_{\rm ss}=2.5\,R_\odot$. The preprint adopts a similar strategy using CR~2154, CR~2157, and CR~2224.

The methodology is therefore equivalent, and the PFSS analysis in the preprint is directly comparable to --- and conceptually derived from --- the approach presented in 2020. A direct discussion of similarities and differences would be beneficial for readers.

\subsection{Cascade Development}
FLUKA fully describes the 3D evolution of electromagnetic and hadronic cascades in the solar atmosphere, enabling predictions for the 511~keV annihilation line, the 2.2~MeV neutron-capture line, and low-energy components. These features cannot be reproduced by the simplified treatment in CRPropa.

\subsection{Primary Spectra and Secondary-Yield Modelling}

While the arXiv paper focuses only on gamma rays, the 2020 study computes a full suite of secondaries, including neutrinos, which enables broader comparisons with multi-messenger constraints.

In the 2020 study, yields were computed for protons, helium, and electrons/positrons using AMS--02 spectra, leading to realistic relative contributions.  
The preprint models only protons and introduces heavy ions via a simple multiplicative factor, neglecting the electron/positron channel altogether. This approximation misses species-dependent effects highlighted in the 2020 analysis.

It is also notable that the preprint cites recent work such as Li et al.\ (Chinese Physics C 48 (4), 045101, 2024), while not citing nor discussing the more comprehensive  multi-species treatment presented in the 2020 study. Given the direct relevance of multi-component primary spectra to the physics of solar gamma-ray production, a comparison with Mazziotta et al.~(2020) would provide important context for assessing the reliability and completeness of the secondary-yield calculations presented in the preprint.

\section{Inverse Compton emission}

A further omission concerns the PRD article by Mazziotta (Phys.\ Rev.\ D~111, 123011 (2025), DOI: 10.1103/zs82-fktf)~\cite{Mazziotta:2025azf}, which presents the first full 3D Monte Carlo calculation of inverse Compton (IC) emission from GCR electrons interacting with solar photons in the presence of magnetic and electric fields. This work demonstrates that the IC component is a significant, magnetically modulated contribution to the solar gamma-ray flux, extending over tens of degrees around the Sun.

Since arXiv:2512.01403 focuses on the role of magnetic-field configurations, ignoring the IC component prevents a complete assessment of solar high-energy emission.

\section{Conclusion}
This comment clarifies the connection between the preprint arXiv:2512.01403 and previous studies modelling GCR interactions with the Sun. The results of Phys.\ Rev.\ D~101 (2020) and Phys.\ Rev.\ D~111 (2025) provide essential context --- respectively in the treatment of hadronic cascades and inverse Compton emission --- and should be considered to fully interpret the predictions presented in the preprint.

\end{document}